\documentclass[
    aps,
    prd,
    reprint,
    amsmath,
    amssymb,
    superscriptaddress,
    nofootinbib
]{revtex4-2}
\usepackage{graphicx}
\usepackage{amsmath}
\usepackage{amsfonts}
\usepackage{amssymb}
\usepackage[normalem]{ulem}
\usepackage{color}%
\usepackage{ragged2e}
\usepackage{dcolumn}
\usepackage{comment}
\usepackage{etoolbox}

\usepackage{MnSymbol,wasysym}
\usepackage{braket}
\usepackage{eurosym}
\usepackage{calrsfs}
\usepackage{multirow}
\usepackage{booktabs}
\usepackage{array}
\usepackage[usenames,dvipsnames,svgnames]{xcolor}

\newcommand{\RNum}[1]{\uppercase\expandafter{\romannumeral #1\relax}}
\usepackage[colorlinks=true,linkcolor=blue,urlcolor=blue,filecolor=black,citecolor=red,pdfstartview=FitV,pdftitle={},pdfsubject={},pdfkeywords={},bookmarksopen=true]{hyperref}
\usepackage[title]{appendix}
\makeatletter
\newcommand*{\rom}[1]{\expandafter\@slowromancap\romannumeral #1@}
\makeatother

\begin{document}

\title{Scalar wave scattering by black holes embedded in dark matter halos}

\author{Shi-Qian Hu}
\email[Corresponding author: ]{shiqian.HU@univ-amu.fr}
\affiliation{Aix Marseille Université, Université de Toulon, CNRS, CPT, Marseille, France}
\affiliation{Université Paris-Saclay, CNRS, CEA, Institut de Physique Théorique,
91191 Gif-sur-Yvette, France}

\author{Federico Piazza}

\affiliation{Aix Marseille Université, Université de Toulon, CNRS, CPT, Marseille, France}

\begin{abstract} 
We study the scattering of massless scalar waves by static black holes embedded in dark matter halos. The halo is modelled by means of a Hernquist profile, for which the Einstein equations admit an exact solution. We analyse the scattering using null geodesics, the glory approximation, the partial wave expansion, and the complex angular momentum (CAM) method.
 The geometric analysis shows that the presence of the halo changes the light ring radius weakly but shifts the critical and glory impact parameters. The wave analysis shows that the characteristic black hole interference pattern is preserved, with halo compactness providing the leading correction and the halo scale radius playing a secondary role. For sufficiently compact halos, the environment also induces a sizeable gravitational redshift. 
 Most of the displacement of the interference extrema can be reproduced by a vacuum Schwarzschild solution with an appropriately chosen $\omega M_{\rm BH}$. A non-vanishing residual, however, reveals a genuine halo contribution that cannot be absorbed into such an effective frequency shift. Finally, we perform the CAM reconstruction of the scattering cross section and find that only a few Regge poles are required to achieve high accuracy at large angles.
\end{abstract}

\maketitle
\newpage
\section{Introduction}

The direct detection of gravitational waves from compact binary mergers \cite{LIGOScientific:2016aoc,LIGOScientific:2016vbw,LIGOScientific:2017vwq}, horizon scale imaging of the Galactic Centre \cite{EventHorizonTelescope:2019dse,EventHorizonTelescope:2022wkp}, and complementary electromagnetic observations \cite{Psaltis:2008bb,GRAVITY:2018ofz} have established black holes as precision laboratories for probing strong gravity. The increasing precision of these observations calls for an increasingly accurate and realistic modelling of \emph{black hole environments}. In fact, black holes are seldom isolated vacuum objects. Environmental effects include the presence of a third body perturbing a binary merger~\cite{Kuntz:2021ohi,Kuntz:2022onu}, as well as thin shells, accretion discs, and cosmological fields surrounding a black~hole~\cite{Visser:1992qh,Abramowicz:2011xu,Macedo:2013qea,Herdeiro:2015waa,Destounis:2025tjn}. 
In this paper we focus on spherically symmetric halos of non-relativistic (dark-) matter surrounding a Schwarzschild black hole. 
This model benefits from known exact solutions of the  Einstein equations which relate the halo profile, mass, and scale radius directly to the spacetime metric \cite{Cardoso:2021wlq}. 

Previous studies have shown that environmental sensitivity depends on the observable. A halo can redshift the quasinormal mode spectrum \cite{Cardoso:2021wlq,Pezzella:2024tkf}, while the associated ringdown waveform can remain nearly degenerate with a vacuum signal once the intrinsic parameters are readjusted \cite{Spieksma:2024voy}. Spin and other gravitational phenomena have also been explored in matter supported black hole geometries \cite{Volkov:1998cc,Fernandes:2025osu,Cardoso:2022whc,Kouniatalis:2025itj,Tian:2025uvk,PhysRevD.111.103007,Chakravarti:2025awj,Bhattacharya:2025lvn,Figueiredo:2023gas,Speeney:2024mas}. These results motivate observables with structure beyond a single characteristic frequency. 

Wave scattering provides such a diagnostic. A monochromatic wave incident from infinity is partly absorbed and partly scattered, producing a signal determined by both weak and strong deflection trajectories \cite{Futterman_Handler_Matzner_1988}. Small angle scattering is dominated by waves that remain comparatively far from the black hole, whereas the large angle interference pattern and the backward glory are generated by trajectories that approach the unstable null orbit \cite{Sanchez:1976fcl,Anninos:1992ih,Andersson:2000tf,Matzner:1985rjn,Zhang:1984vt,Batic:2012rm}. Different angular sectors thus sample different radial regions of the geometry.

Scattering by vacuum black holes has been studied extensively for several field spins and background geometries \cite{Dolan:2006vj,Crispino:2009xt,Dolan:2008kf,Dolan:2009zza,Crispino:2009ki,Benone:2014qaa,Macedo:2015qma,Gussmann:2016mkp,Cotaescu:2018etx}. Environmental effects have also been investigated for black holes surrounded by matter distributions \cite{Torres:2022fyf, Heidari:2026swh}. 
A key feature of black hole scattering is its spectral description in terms of quasinormal modes (QNMs) and Regge poles (RPs). These two spectra represent complementary aspects of the same resonant structure: QNMs are obtained by fixing the angular momentum and solving for complex frequencies, whereas RPs are obtained by fixing the frequency and analytically continuing the angular momentum into the complex plane \cite{Dreyer:2003bv, Berti:2009kk,Glampedakis:2003dn,Decanini:2002ha,Folacci:2019cmc}.

Complex angular momentum (CAM) representations of black hole scattering have been developed extensively in vacuum spacetimes, where the large angle cross section can be reconstructed efficiently from the sum of RPs and the background contribution \cite{Folacci:2018sef,Folacci:2019cmc,Folacci:2019vtt}. Environmental effects have also been investigated in black hole spacetimes containing localised matter distributions, where additional RP branches may appear even though the reconstructed scattering observable remains comparatively stable \cite{Torres:2022fyf,Torres:2023nqg}. These results show that the motion of individual poles does not by itself determine the change in the scattering cross section. It remains unknown how this description extends to a smooth and spatially distributed halo described by an exact black hole solution.

In this work, we study massless scalar waves propagating on black holes surrounded by dark matter halos. The null geodesics and deflection function determine the light ring, the critical impact parameter separating capture from scattering, and the glory impact parameter. The differential scattering cross section is obtained from the exact partial wave expansion \cite{Andersson:1995vi,Glampedakis:2001cx,Sanchez:1977vz}. We then analytically continue the scattering matrix to determine the RPs and apply the CAM reconstruction \cite{Andersson:1994rk,Decanini:2002ha,Glampedakis:2003dn,Folacci:2018sef,Folacci:2019cmc,Folacci:2019vtt}. These complementary descriptions capture the same halo induced physics from three perspectives: classical trajectories, wave scattering, and the resonance spectrum.

One useful parameter characterizing the halo is \emph{compactness} $z = M/a_0$, with $M$ and $a_0$ the halo mass and size, respectively. In geometrized units, $z$ indicates how much the matter distribution contributes to the geometry in the strong gravity region close to the black hole.
Our results identify two distinct environmental effects. First, the dominant displacement of the interference extrema is consistent with an effective gravitational redshift, although the halo cross section cannot be reproduced completely by tuning the Schwarzschild frequency alone. Second, the RPs undergo a coherent displacement rather than the branch reorganisation associated with a narrow localised perturbation. Within the CAM representation, we obtain systematic convergence to the partial wave result as the number of RPs is increased. The first six poles can accurately reconstruct the scattering cross section at large angles.

Appreciable differences are obtained, with respect to the vacuum case, only with high compactness values, $z\simeq 10^{-2} - 10^{-1}$. Conventional galactic dark matter halos are expected to be substantially more dilute. A galaxy like the Milky Way has $z\simeq 10^{-6}$. However, the compactness can increase in the region closer to the black hole. As summarized, e.g. in \cite{Berti:2025hly, Jiao:2023mw}, the black hole at the centre of the Milky Way is surrounded by stars and dark matter which, within a radius of $a_0\sim 20 {\rm kpc}$, form a halo of compactness $z\sim 10^{-4}$. Consistent with the compactness dependence found in our numerical results, the scattering displacement associated with such a dilute halo is therefore expected to be much smaller than in our benchmark configurations. The high $z$ configurations considered here should be interpreted as theoretical benchmarks for the underlying physical mechanism rather than direct models of the Milky Way halo. More concentrated structures, such as dark matter spikes or other compact matter distributions, may enhance the gravitational potential in the vicinity of the black hole, but their relation to the global compactness parameter $z$ is model dependent \cite{Gondolo:1999ef,Hertzberg:2019exb,Oguri:2022fir,Bertone:2024rxe}.

Regardless of how realistic the compactness parameters may be in astrophysical settings, studying the effect of the halo is also of theoretical interest. In particular, for the scalar probe considered here, we expect that the halo induced effects can be matched from the underlying GR description onto appropriate finite size response coefficients in the EFT of compact objects, building on the worldline EFT framework of Goldberger and Rothstein~\cite{Goldberger:2004jt} (see~\cite{Huang:2018pbu,Kuntz:2019zef} for more specific applications to a scalar field and~\cite{Ivanov:2024sds} for a more recent scattering based matching analysis.) 

This paper is organised as follows. In Sec.~\ref{sec: bh}, we introduce the black hole geometries and halo profiles. Section~\ref{sec: geodesics} presents the null geodesic analysis and glory approximation. In Sec.~\ref{sec: pwe}, we compute the scalar scattering cross section with a partial wave expansion. Section~\ref{sec: rp} develops the complex angular momentum representation, determines the RPs and tests CAM method. We summarise the results in Sec.~\ref{sec:conclusion}. Geometrised units with $G=c=1$ are used throughout.

\section{Black hole with dark matter halo}\label{sec: bh}
In this section, we review the key equations governing black hole solutions embedded in dark matter halos. The fully relativistic geometry describes a black hole in the near region while reproducing a dark matter distribution on large scales \cite{Figueiredo:2023gas}.
In this framework, the background of the system is effectively described by an anisotropic stress-energy tensor 
\begin{equation}
    T^{\mu}_{\nu} = \text{diag}(-\rho, P_r, P_t, P_t),
\end{equation}
where $\rho$ is the energy density, $P_r$ is the radial pressure, and $P_t$ is the tangential pressure.
We assume spherical symmetry, so all components of the stress energy tensor depend only on $r$. The black hole solution is obtained in closed analytic form and is described by the metric \cite{Cardoso:2021wlq}
\begin{equation}\label{eq:metric}
ds^2 = -f(r) dt^2 + \frac{dr^2}{1 - 2m(r)/r} + r^2 (d\theta^2 + \sin^2 \theta d\varphi^2),
\end{equation}
where $m(r)$ denotes the mass function. With the vanishing radial pressure $P_r = 0$, the Einstein equations reduce to 
\begin{align}\label{fieldeqn}
m'(r) &=  4\pi r^2 \rho,  \nonumber \\ 
\frac{f'(r)}{f(r)} &= \frac{2m(r)}{r(r-2m(r))}, \\
2P_t &= \frac{ m(r)}{r-2m(r)}\rho. \nonumber
\end{align}
Equations \eqref{fieldeqn} determine the background functions $f(r)$ and $m(r)$ for a specified density profile $\rho(r)$.
We consider a class of parametric density distributions commonly used to model cold dark matter halos in different galactic environments. These profiles can be written in the unified form \cite{Figueiredo:2023gas,Pezzella:2024tkf}
\begin{equation}
\rho(r) = \rho_0 \left(\frac{r}{a_0}\right)^{-\gamma}
\left[1 + \left(\frac{r}{a_0}\right)^{\alpha}\right]^{(\gamma-\beta)/\alpha},
\end{equation}
where $a_0$ is the characteristic halo scale radius and $\rho_0$ is the density normalisation.
The parameters $\gamma$ and $\beta$ control the inner and outer slopes of the density profile, respectively, while $\alpha$ determines the sharpness of the transition between these asymptotic regimes \cite{Graham:2005xx}.
Two models have been studied extensively and tested against observational data and $N$-body simulations: the Hernquist profile, obtained by setting $(\alpha,\beta,\gamma)=(1,4,1)$ \cite{Hernquist:1990be}, and the Navarro--Frenk--White (NFW) profile, corresponding to $(\alpha,\beta,\gamma)=(1,3,1)$ \cite{Navarro:1996gj}. Because the total mass of the NFW profile diverges, a finite outer cutoff radius is required in practical implementations.
\begin{figure*}[t]
    \centering
    \includegraphics[width=0.85\textwidth]{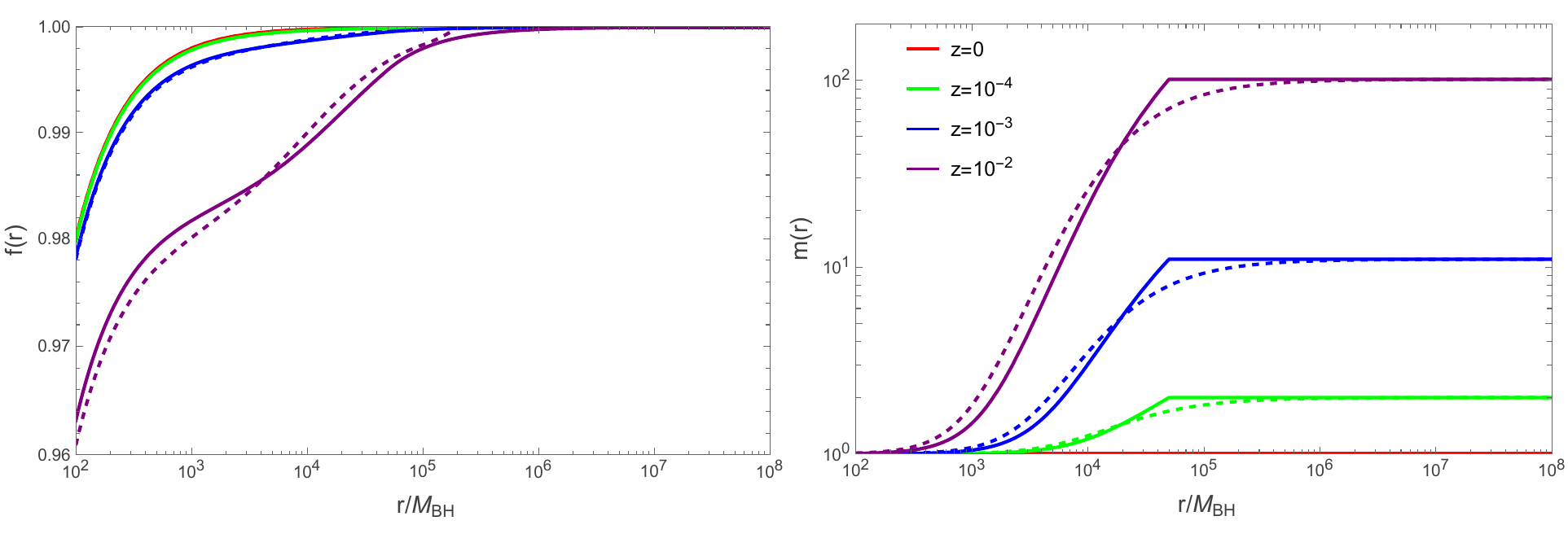}
    \caption{ Metric and mass functions for black holes embedded in dark matter halos. The left panel shows the metric function $f(r)$, while the right panel shows the mass function $m(r)$, for different values of the halo compactness parameter $z$: $10^{-4}$ (green), $10^{-3}$ (blue), $10^{-2}$ (purple). For $z=0$ (red), we recover the Schwarzschild solution. Solid curves denote the NFW profile and dashed curves denote the Hernquist profile. We fix $a_0=10^4 $ and $M_{\rm BH}=1$.}
    \label{fig:mfplot}
\end{figure*}
Fig.~\ref{fig:mfplot} shows the metric and mass functions for both profiles at several values of the compactness parameter defined in \eqref{compactness}. The NFW cutoff is chosen so that the total halo masses coincide. 
The two profiles display qualitatively similar behaviour at the resolution considered.
We therefore focus on the Hernquist profile in the remainder of the paper.
As opposed to other models for which the metric must be integrated numerically (e.g. \cite{Figueiredo:2023gas}), the Hernquist profile enjoys an exact closed form for the mass function~\cite{Cardoso:2021wlq, Pezzella:2024tkf},
\begin{equation} \label{massfunction}
  m(r) = M_{\text{BH}} + \frac{ M r^2}{(a_0 + r)^2}  \left(1 - \frac{2M_{\text{BH}}}{r} \right)^2 \, .
\end{equation}  
The corresponding lapse function appearing in \eqref{eq:metric} is
\begin{equation}
    f(r)=\left(1-\frac{2M_{\rm BH}}{r}\right)e^{\Gamma(r)},
\label{eq:hernquist_lapse}
\end{equation}
with
\begin{equation}
\begin{aligned}
    \Gamma(r)&=-\pi\sqrt{\frac{M}{\xi}}
+2\sqrt{\frac{M}{\xi}}
\arctan\!\left(\frac{r+a_0-M}{\sqrt{M\xi}}\right),\\
\qquad
&\xi\equiv 2a_0-M+4M_{\rm BH}>0.
\label{eq:hernquist_gamma}
\end{aligned}
\end{equation}
This geometry describes a central black hole surrounded by a Hernquist dark matter halo. It represents a near horizon deformation of the Newtonian Hernquist profile and reduces to it in the dilute halo limit  $M_{\rm BH}/a_0\ll1$ away from the horizon. At large radii,
\begin{equation}
f(r)=1-\frac{2M_{\rm ADM}}{r}+\mathcal O(r^{-2}),
\qquad
M_{\rm ADM}=M_{\rm BH}+M,
\end{equation}
showing that the asymptotic gravitational field is determined by the total ADM mass. The Schwarzschild geometry is recovered for $M=0$.
For convenience, we define the dimensionless halo compactness parameter
\begin{equation}\label{compactness}
    z \equiv \frac{M}{a_0}.
\end{equation}
The resulting solutions for $f(r)$ and $m(r)$ determine the geodesic structure of the spacetime, including characteristic quantities such as the innermost stable circular orbit and the light rings \cite{Fonseca:2025ehf}. These quantities encode essential features of the background geometry and are closely related to observable phenomena such as wave scattering, thereby providing a natural starting point for investigating how the surrounding dark matter environment affects the scattering cross section. 
\section{Wave scattering: Geodesic analysis and the glory approximation} \label{sec: geodesics}
In this section, we investigate how a dark matter halo modifies wave scattering by a black hole. We first study the null geodesics and the associated glory approximation, before turning to the full wave calculation in Sec.~IV.

\subsection{Geodesic analysis}
The background solution determines the geodesic motion of both massive and massless particles. Stationarity and spherical symmetry imply conservation of the energy $E$ and total angular momentum $L$. Each orbit may be chosen to lie in the equatorial plane. The radial geodesic equation can be written as
\begin{equation}
\frac{f(r)}{1-2m(r)/r} \dot r^2 + V_{\rm eff}(r) = E^2 ,
\label{eq:radial_motion}
\end{equation}
where $V_{\rm eff}$ is the effective potential
\begin{equation}
V_{\rm eff}(r) = f(r)\left( \kappa + \frac{L^2}{r^2} \right)\, .
\label{eq:Veff}
\end{equation}
In the above, $\kappa=1$ for timelike geodesics and $\kappa=0$ for null geodesics. 
Circular geodesics correspond to stationary points of the effective potential. 
For timelike circular geodesics $\kappa=1$ at $r=r_p$, the conditions
\begin{equation}
    V_{\rm eff}(r_p)=E^2, 
\qquad 
V'_{\rm eff}(r_p)=0
\end{equation}
yield
\begin{equation}
E 
=
\left[ \frac{r - 2m(r)}{r - 3m(r)} \, f(r) \right]^{1/2}_{r=r_p}, 
L 
= \left[ \frac{ r^2 m(r)}{r - 3m(r)} \right]^{1/2}_{r=r_p}.
\label{eq:EL}
\end{equation}
The light ring radius is therefore determined implicitly by $r_{\rm LR}=3m(r_{\rm LR})$, which has the same form as the Schwarzschild condition when expressed in terms of the mass function.
For a black hole surrounded by matter, however, the mass function $m(r)$ also contains the halo contribution, shifting the light ring relative to its vacuum value. 
\begin{figure}[t]
\centering
\includegraphics[width=0.9\columnwidth,    height=0.58\columnwidth,
keepaspectratio]{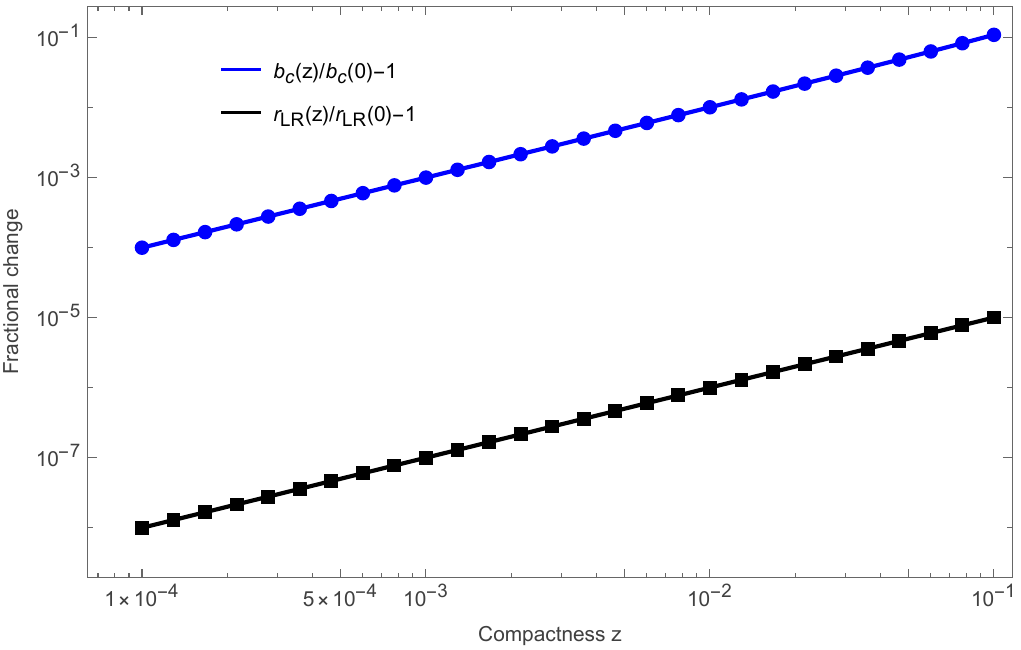}
\caption{ Fractional changes in the critical impact parameter, $b_c(z)/b_c(0)-1$, and in the light ring radius, $r_{\text{LR}}(z)/r_{\text{LR}}(0)-1$, as functions of the halo compactness parameter $z$ for a black hole surrounded by a Hernquist dark matter halo, where we fix $a_0=10^4$ and $M_{\rm BH}=1$. The data points are sampled in the range $10^{-4}\leq z \leq 10^{-1}$, logarithmically spaced in $z$.}
\label{fig:bc_rph_vs_z}
\end{figure}
As shown in Fig.~\ref{fig:bc_rph_vs_z}, both the light ring $r_{\rm LR}$ and the critical impact parameter $b_c$ increase monotonically with the halo compactness parameter $z$. The fractional change in the light ring position remains very small throughout the range considered, while the critical impact parameter exhibits a larger increase.
This shift modifies the boundary between photon capture and scattering trajectories. To examine the corresponding null geodesics in the equatorial plane, we set $\kappa=0$ in \eqref{eq:radial_motion}. Together with $\dot\phi=L/r^2$, this gives the orbit equation
\begin{equation}
\frac{d\phi}{dr}=
\frac{b}{r^2
\sqrt{
\left(1-\frac{2m(r)}{r}\right)
\left(\frac{1}{f(r)}-\frac{b^2}{r^2}\right)
}
}
\label{eq:orbit_equation}
\end{equation}
The total deflection angle for a trajectory with impact parameter $b$ is then obtained by integrating the orbit equation,
\begin{equation}
\Theta(b)=
2\int_{r_0}^{\infty}
\frac{b}
{r^2\sqrt{
\left(1-\frac{2m(r)}{r}\right)
\left(\frac{1}{f(r)}-\frac{b^2}{r^2}\right)}}dr-\pi 
\label{eq:deflection_angle}
\end{equation}
where $r_0$ denotes the distance of closest approach.
From \eqref{eq:deflection_angle}, the deflection angle diverges logarithmically when the impact parameter approaches its critical value. In this limit, the photon trajectories display the characteristic orbiting behaviour associated with the unstable circular orbit at the light ring. The critical impact parameter is affected both by the displacement of the light ring and the halo induced modification of the lapse evaluated at the light ring.

\begin{figure}[t]
\centering
\includegraphics[width=0.95\columnwidth,
    height=0.58\columnwidth,keepaspectratio]{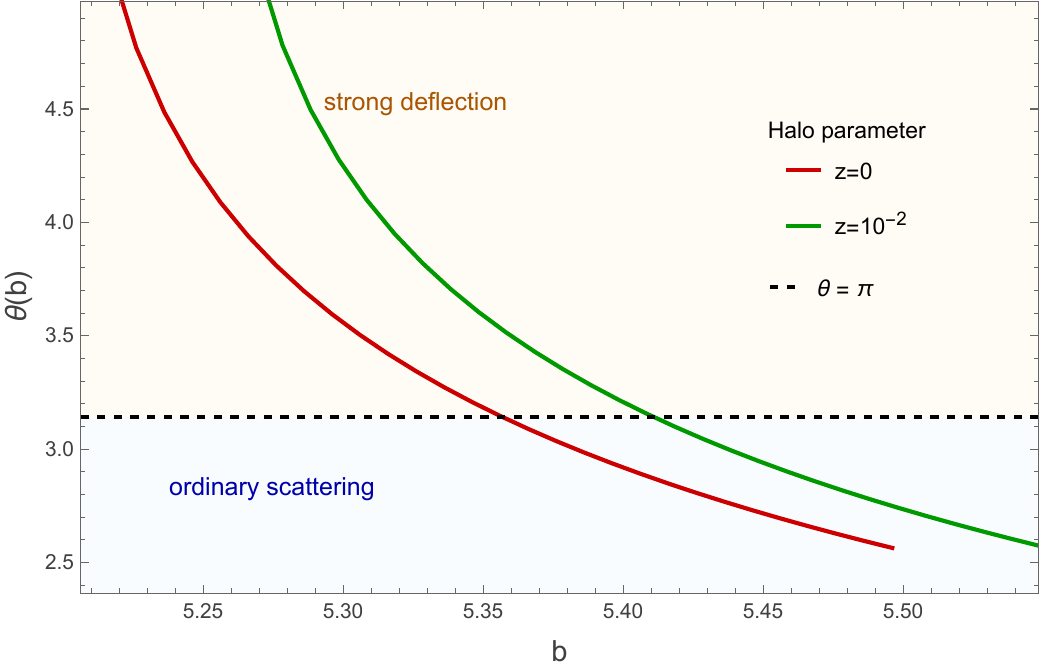}
\caption{
Deflection function $\Theta(b)$ for different values of compactness $z$ in the Hernquist halo model. The dashed horizontal line marks $\theta=\pi$. Its intersection with each curve defines the corresponding glory impact parameter $b_g(z)$.}
\label{fig:deflection_halo}
\end{figure}

\begin{figure}[t]
\centering
\includegraphics[width=0.7\columnwidth]{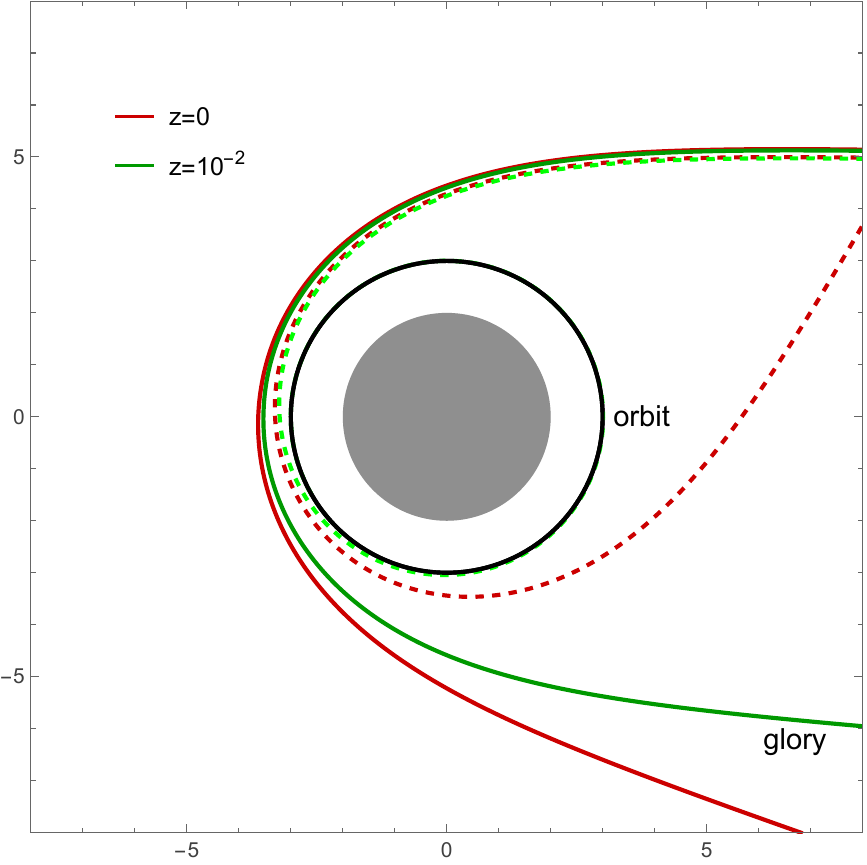}
\caption{ 
Representative null geodesics in the Hernquist halo model with $z=10^{-2}$ (green) and in the Schwarzschild background (red). The shaded disk denotes the black hole region, while the black circle marks the light ring. The solid green trajectory is plotted at the glory impact parameter $b_g$, and the dashed green trajectory is plotted at the halo critical impact parameter $b_c(z=10^{-2})$. The red curves show Schwarzschild geodesics with the same impact parameters.}
\label{fig:geodesics_halo}
\end{figure}

Fig.~\ref{fig:deflection_halo} shows the deflection angle $\Theta(b)$, defined in \eqref{eq:deflection_angle}, as a function of the impact parameter $b$ for different values of $z$. Increasing $z$ shifts the curves upwards and to the right, so a larger impact parameter is required to produce the same deflection. Both the critical point associated with orbiting and the glory point defined by $\Theta(b_g)=\pi$ move to larger impact parameters as the halo compactness increases. Thus, a dark matter halo shifts the onset of orbiting and the condition for backward scattering, modifying the near critical null geodesic structure.
The impact parameter characterises the asymptotic transverse offset of a photon incident from infinity, fixing the same numerical value of $b$ therefore permits a direct comparison of trajectories in different backgrounds.
Fig.~\ref{fig:geodesics_halo} shows null geodesics for two fixed values of the impact parameter $b$ in the Schwarzschild background and in the halo background. We choose $b=b_c$ and $b=b_g$ as corresponding to the orbiting and glory trajectories in the spacetime of the Hernquist profile with $z=10^{-2}$. The shaded grey disk represents the black hole region, while the black circle marks the light ring. In the halo background, the dashed green trajectory with $b=b_c$ undergoes orbiting near the light ring, and the solid green trajectory with $b=b_g$ is backward scattered, corresponding to a deflection angle $\theta=\pi$. 
As expected, without the contribution of the halo mass (i.e. on a Schwarzschild solution), the same values of $b$ give rise to null trajectories (red curves in Fig.~\ref{fig:geodesics_halo}) that undergo a smaller deflection in the Schwarzschild background.
\subsection{Glory approximation}
\label{sec:glory_approximation}

\begin{figure}[t]
\centering
\includegraphics[width=0.95\columnwidth]{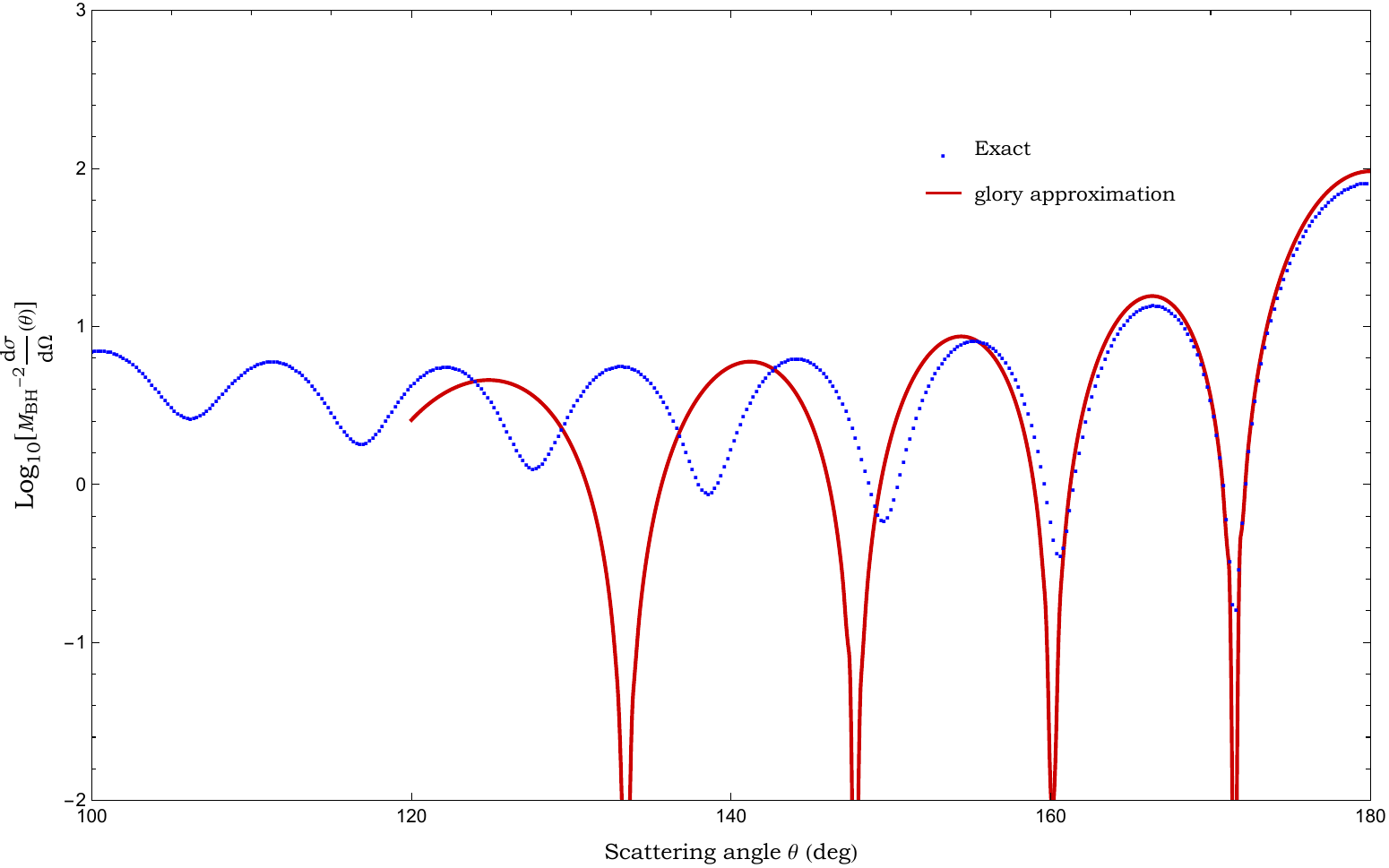}
\caption{ Differential scattering cross section as a function of the scattering angle $\theta$ for the Hernquist profile with $(z,a_0/M_{\rm BH})=(10^{-2},10^2)$ at $\omega M_{\rm BH}=3$. The blue dashed curve shows the exact partial wave expansion, while the red solid curve corresponds to the glory approximation.}
\label{fig:glory}
\end{figure}

The light ring plays a central role not only in the geodesic structure of the spacetime, but also in the strong field scattering of waves propagating on this background. The differential scattering cross section near the backward direction is governed by trajectories with impact parameters close to the glory value, determined by $\Theta(b)=\pi$. The large angle scattering behaviour can therefore be inferred, at the semiclassical level, from the corresponding null geodesic structure. For scalar field scattering, the differential cross section in the glory approximation is given by \cite{Matzner:1985rjn, DeWitt-Morette:1984hsq, Zhang:1984vt, Anninos:1992ih} 
\begin{equation}
    \left.\frac{d\sigma}{d\Omega}\right|_{\mathrm{glory}}
    \simeq
    2\pi \omega b_g^2
    \left|\frac{db}{d\theta}\right|_{\theta=\pi}
    \left[
    J_0\!\left(\omega b_g \sin\theta\right)
    \right]^2 ,
    \label{eq:gloryapprox}
\end{equation} 
where $\omega$ is the wave frequency, $b_g$ is the glory impact parameter, and $J_0$ is the Bessel function of the first kind. The derivative factor is determined by the local behaviour of the deflection function,
\begin{equation}
    \left|
    \frac{db}{d\theta}
    \right|_{\theta=\pi}
    =
    \left|
    \frac{1}{d\Theta/db}
    \right|_{b=b_g}.
\end{equation}
Fig.~\ref{fig:glory} compares this semiclassical prediction with the full numerical partial wave result for the same Hernquist configuration. The moderately high frequency $\omega M_{\rm BH}=3$ is chosen because the semiclassical glory approximation becomes more accurate in the short wavelength regime.
Near the backward direction, the interference pattern is well described by the glory approximation. The halo modifies the deflection function and consequently shifts the glory impact parameter and its derivative factor, changing both the angular positions and amplitudes of the backward fringes. This provides a direct semiclassical connection between the modified null geodesics and the observable scattering pattern. However, to obtain the full wave scattering cross section beyond the semiclassical approximation, it is necessary to solve the scattering problem exactly. We therefore turn to the partial wave expansion method, which allows us to compute the full scattering cross section and to study how the modifications of the background geometry are imprinted on the observable wave scattering signal.

\section{Scalar wave scattering} \label{sec: pwe}

\subsection{Field equation and scattering amplitude}

We consider a massless scalar field propagating on the fixed black hole and halo background. Perturbations of the matter distribution are neglected, so the field satisfies the minimally coupled Klein--Gordon equation
\begin{equation}\label{eq:KG_scalar}
    \Box\Phi
    =\frac{1}{\sqrt{-g}}\partial_{\mu}
    \left(\sqrt{-g}\,g^{\mu\nu}\partial_{\nu}\Phi\right)=0,
\end{equation}
where $g$ is the determinant of $g_{\mu\nu}$. We choose the incident direction as the $z$ axis. Axisymmetry then selects the $m=0$ sector, and the field can be written as
\begin{equation}\label{eq:scalar_decomposition}
    \Phi(t,r,\theta)
    =\frac{e^{-i\omega t}}{r}
    \sum_{\ell=0}^{\infty}(2\ell+1)
    P_{\ell}(\cos\theta)\,\phi_{\omega\ell}(r).
\end{equation}

The radial wave equation takes a Schr\"odinger form when expressed in the tortoise coordinate
\begin{equation}\label{eq:tortoise}
    \frac{dr_*}{dr}
    =\left[f(r)\left(1-\frac{2m(r)}{r}\right)\right]^{-1/2}.
\end{equation}
The horizon and spatial infinity are mapped to $r_*\to-\infty$ and $r_*\to+\infty$, respectively. For a scalar field, the radial equation is
\begin{equation}\label{radialeqn}
    \left[
    \frac{d^2}{dr_*^2}+\omega^2-V_{\ell}(r)
    \right]\phi_{\omega\ell}(r)=0,
\end{equation}
\begin{equation}\label{eq:dbhpoten2}
    V_{\ell}(r)
    =f(r)\left[
    \frac{\ell(\ell+1)}{r^2}
    +\frac{2m(r)}{r^3}
    -\frac{m'(r)}{r^2}
    \right].
\end{equation}
The halo enters through both $f(r)$ and $m(r)$ and therefore modifies the height, location, and asymptotic tail of the effective potential.

We use the solution that is purely ingoing at the horizon and contains incoming and outgoing components at infinity:
\begin{equation}\label{bc_in}
    \phi_{\omega\ell}^{\rm in}(r_*)\sim
    \begin{cases}
       e^{-i\omega r_*}, & r_*\to-\infty,\\[2mm]
       A_{\ell}^{\rm in}e^{-i\omega r_*}
       +A_{\ell}^{\rm out}e^{+i\omega r_*},
       & r_*\to+\infty.
    \end{cases}
\end{equation}
The partial wave matrix is
\begin{equation}\label{Matrix_S}
    S_{\ell}(\omega)
    =(-1)^{\ell+1}
    \frac{A_{\ell}^{\rm out}(\omega)}
         {A_{\ell}^{\rm in}(\omega)}.
\end{equation}
After subtracting the incident wave, the scattering amplitude and differential cross section are
\begin{align}
    f(\theta)
    &=\frac{1}{2i\omega}
      \sum_{\ell=0}^{\infty}(2\ell+1)
      \left[S_{\ell}(\omega)-1\right]
      P_{\ell}(\cos\theta),
      \label{eq:scatampt}\\
    \frac{d\sigma}{d\Omega}
    &=\left|f(\theta)\right|^2.
    \label{eq:diff_cross_section}
\end{align}
These expressions provide the exact wave description used below and complement the geodesic and glory approximations.

\subsection{Numerical implementation}

We integrate \eqref{radialeqn} with the purely ingoing boundary condition at the horizon and, in the far zone, match the numerical solution to incoming and outgoing waves whose long-range logarithmic phase is determined by the total ADM mass, $M_{\rm ADM}=M_{\rm BH}+M$. The long-range gravitational interaction causes the partial wave series to converge slowly, especially in the forward direction \cite{Andersson:2000tf,Taylor1974}. To accelerate convergence, we use the iterative Yennie--Ravenhall--Wilson (YRW) series reduction \cite{PhysRev.95.500}. Defining $\widetilde a_{\ell}^{(0)}=(2\ell+1)[S_{\ell}(\omega)-1]$, the amplitude after $N$ reductions is
\begin{equation}
f(\theta)=\frac{1}{2i\omega(1-\cos\theta)^{N}}
\sum_{\ell=0}^{\infty}\widetilde a_{\ell}^{(N)}P_{\ell}(\cos\theta),
\label{eq:yrw_amp}
\end{equation}
where
\begin{equation}
\widetilde a_{\ell}^{(n)}
=
\widetilde a_{\ell}^{(n-1)}
-\frac{\ell+1}{2\ell+3}\widetilde a_{\ell+1}^{(n-1)}
-\frac{\ell}{2\ell-1}\widetilde a_{\ell-1}^{(n-1)},
 n=1,\ldots,N,
\label{eq:yrw_recursion}
\end{equation}
with $\widetilde a_{-1}^{(n)}=0$. At any fixed angle $\theta\neq0$, this transformation leaves the infinite partial wave amplitude unchanged. For coefficients with a smooth large-$\ell$ asymptotic expansion, each iteration cancels the leading tail terms and typically improves their falloff by two powers of $\ell$ \cite{Folacci:2019cmc}. The explicit factor $(1-\cos\theta)^{-N}$ makes the exact forward limit numerically delicate. We compute the $S$ matrix through $\ell=600$, retain modes through $\ell=400$ in the final reduced sum, and use $N=3$.

\subsection{Scattering results}

We set $M_{\rm BH}=1$ and compare four representative Hernquist halo configurations:
\begin{equation}\label{eq:benchmark_cases}
    \begin{aligned}
    \text{Case 1}:&\quad (z,a_0)=(10^{-1},10^2),\\
    \text{Case 2}:&\quad (z,a_0)=(10^{-2},10^2),\\
    \text{Case 3}:&\quad (z,a_0)=(10^{-2},10^4),\\
    \text{Case 4}:&\quad (z,a_0)=(10^{-3},10^4).
    \end{aligned}
\end{equation}
These configurations allow comparisons at fixed scale radius, fixed compactness, and fixed total halo mass. 
\begin{figure}[t]
\centering
    \includegraphics[width=\columnwidth]{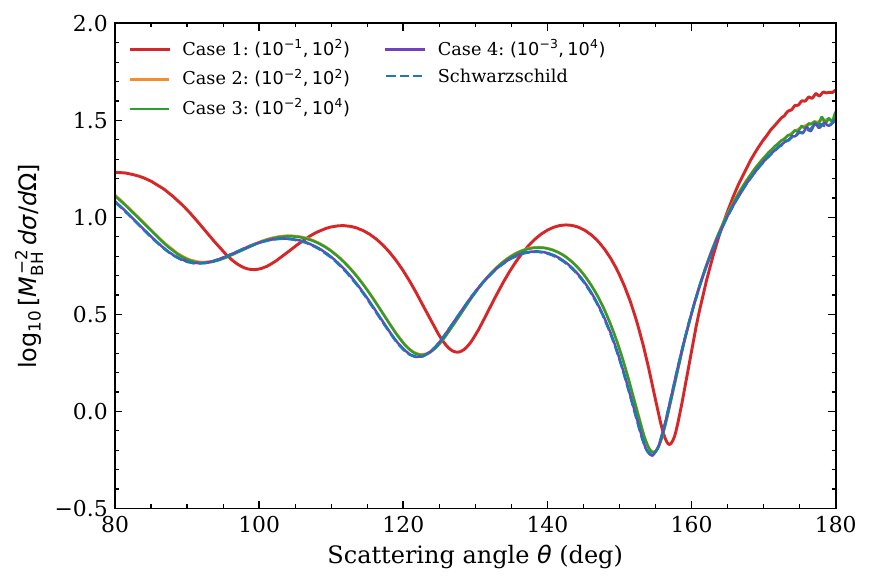}
\caption{ Scalar differential scattering cross sections computed from the partial wave expansion at $M_{\rm BH}=1$ and $\omega=1$. The solid curves denote the four Hernquist configurations defined in \eqref{eq:benchmark_cases}, while the dashed curve denotes Schwarzschild.}
\label{fig:glory2}
\end{figure}
Fig.~\ref{fig:glory2} shows that the deviation from Schwarzschild is not determined by the total halo or ADM mass. Cases~1 and 4 have identical $M$ and $M_{\rm ADM}$, but the more compact Case~1 with $z=10^{-1}$ produces the largest displacement of the interference pattern, whereas Case~4 is the closest to the vacuum result. 
 Cases~2 and 3, which share $z=10^{-2}$, but have different scale radii $a_0$ remain close to one another. 
 Cases~1 and 2 compare different compactnesses at fixed $a_0=10^2$, while Cases~3 and 4 provide the corresponding comparison at fixed $a_0=10^4$. We found that increasing compactness mainly displaces the interference extrema, with smaller changes in their amplitudes. We also concluded that compactness is the leading control parameter, while the scale radius provides a secondary correction.

The displaced minima and maxima in Fig.~\ref{fig:glory2} show that the halo modifies the angular interference pattern rather than producing a uniform enhancement or suppression. We test how much of this displacement is degenerate with a gravitational redshift by comparing the Case~1 scattering cross section at $\omega=1$ with Schwarzschild scattering cross sections evaluated at an effective frequency. Writing $\sigma(\theta)=d\sigma/d\Omega$, the redshift constrained reconstruction is
$\sigma_{\rm rec}(\theta,\omega)
=(\omega^2/\omega_0^2)\sigma_{\rm Schw}(\theta, \omega)$, with $\omega_0=1$. The factor $\omega^2/\omega_0^2$ removes the explicit $1/\omega^2$ prefactor of the differential scattering cross section in \eqref{eq:diff_cross_section}. Guided by the leading environmental redshift estimate of Ref.~\cite{Berti:2025hly}, we scan $0.8\leq\omega\leq1.3$ and minimise
\begin{equation}
\label{eq:redshift_error}
\mathcal E(\omega)=
\left[
\frac{\displaystyle\int_{80^\circ}^{180^\circ}
\left[\omega^2\sigma_{\rm Schw}(\theta,\omega)-\sigma_{\rm H}(\theta,1)\right]^2
\sin\theta\,d\theta}
{\displaystyle\int_{80^\circ}^{180^\circ}
\sigma_{\rm H}^2(\theta,1)\sin\theta\,d\theta}
\right]^{1/2}.
\end{equation}
The minimum occurs at $\omega_{\rm eff}\simeq1.12$, for which $\mathcal E=9.51\%$. 

\begin{figure}[t]
\centering
    \includegraphics[width=\columnwidth]{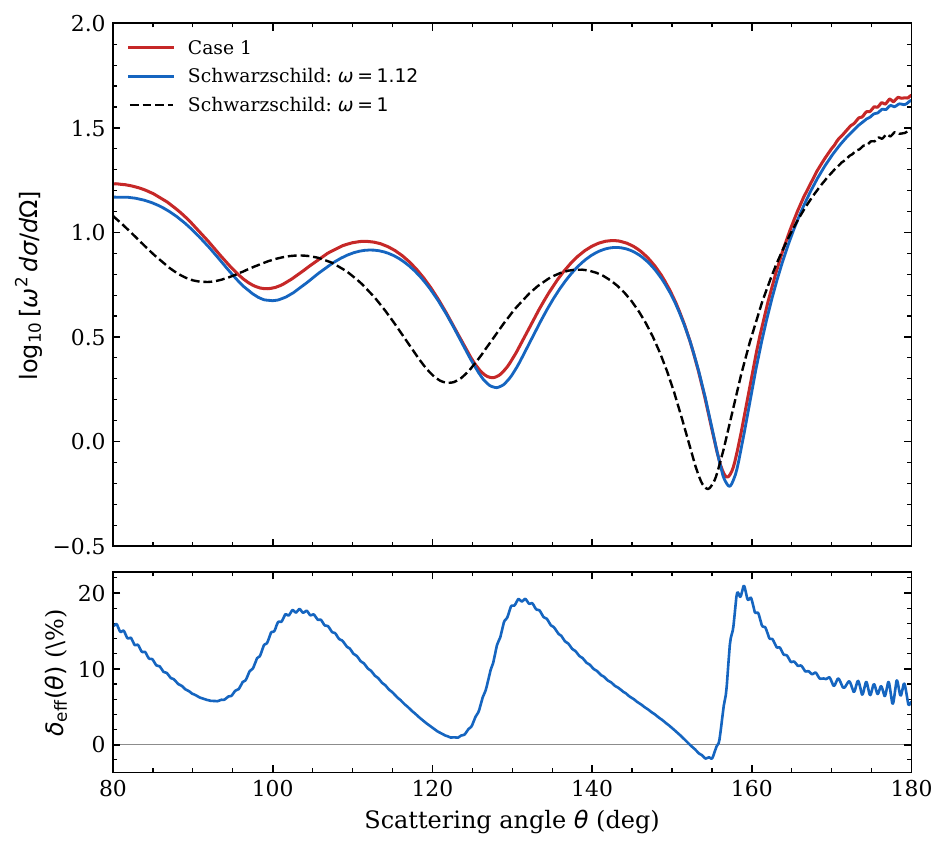}
\caption{ Redshift constrained reconstruction of the Case~1 scattering cross section. The upper panel compares Case~1 at $\omega=1$ (red) with Schwarzschild at $\omega=1$ (dashed) and with the best fit redshift reconstruction at $\omega_{\rm eff}=1.12$ (blue). The lower panel shows the signed pointwise relative residual $\delta_{\rm eff}(\theta)=100\,
\frac{\sigma_{\rm H}(\theta,1)-\omega_{\rm eff}^{2}\sigma_{\rm Schw}(\theta,\omega_{\rm eff})}{\omega_{\rm eff}^{2}\sigma_{\rm Schw}(\theta,\omega_{\rm eff})}$ for the blue curve.}
\label{fig:redshift_reconstruction}
\end{figure}

Fig.~\ref{fig:redshift_reconstruction} compares Case~1 at fixed $\omega=1$ with the Schwarzschild results at $\omega=1$ and $\omega_{\rm eff}=1.12$. The frequency correction reproduces most of the displacement of the interference extrema but not the full amplitude and angular dependence. For $z=10^{-1}$, Eq.~(2.88) of Ref.~\cite{Berti:2025hly} gives $\delta_n\simeq-z+0.17z^2=-0.0983$, corresponding to a compensating frequency $\omega/(1+\delta_n)=1.109$. This lies within approximately $0.9\%$ of the fitted value. Since the quoted relation was derived for axial $\ell=2$ quasinormal modes rather than scalar scattering, the comparison should be regarded as suggestive evidence for a common leading redshift mechanism, not as a direct quantitative test of that formula.

\section{Regge poles and CAM reconstruction}\label{sec: rp}

\subsection{CAM representation}

The partial wave expansion describes the scattering amplitude as a sum over integer angular momenta. In the complex angular momentum (CAM) formulation, the partial wave
$S$ matrix is analytically continued from integer angular momentum
$\ell$ to the complex variable $\lambda=\ell+1/2$, yielding
$S_{\lambda-1/2}(\omega)$
\cite{Folacci:2018sef,Folacci:2019cmc,Folacci:2019vtt}.
At a fixed real frequency, the RPs are the
simple poles of this analytically continued $S$ matrix. In terms of the
asymptotic amplitudes defined in \eqref{bc_in}, they satisfy
\begin{equation}
\label{eq:RP_condition}
A_{\lambda_n-1/2}^{\rm in}(\omega)=0,
\qquad
\omega\in\mathbb{R},
\quad
\lambda_n\in\mathbb{C}.
\end{equation}
In the following, $\lambda_n$, $n=0,1 \dots$, denotes the RPs in the first quadrant, ordered by increasing imaginary part. With the analytic continuation of \eqref{Matrix_S}, our convention for the scattering matrix is
\begin{equation}\label{eq:CAM_Smatrix}
S_{\lambda-1/2}(\omega)=e^{i\pi(\lambda+1/2)}\frac{A_{\lambda-1/2}^{\rm out}(\omega)}     {A_{\lambda-1/2}^{\rm in}(\omega)}.
\end{equation}
The residue associated with a simple pole is therefore
\begin{equation}\label{eq:RP_residue_definition}
\begin{aligned}
r_n(\omega)&=e^{i\pi[\lambda_n(\omega)+1/2]}\left.\frac{A_{\lambda-1/2}^{\rm out}(\omega)}     {\partial_\lambda A_{\lambda-1/2}^{\rm in}(\omega)}\right|_{\lambda=\lambda_n(\omega)} 
\end{aligned}
\end{equation}
These poles determine the resonant surface wave contributions entering
the complex angular momentum representation of the scattering amplitude \cite{Andersson:1994rk,Decanini:2002ha,Glampedakis:2003dn}. Semiclassically, the RPs describe surface waves trapped near the unstable null orbit, which decay as they propagate around the black hole. A Sommerfeld--Watson transformation separates the scattering amplitude into a background contribution and a sum over RPs \cite{Folacci:2019cmc,Folacci:2019vtt},
\begin{equation}\label{eq:CAM_decomposition}
    f(\omega,\theta)
    =f^{\rm B}(\omega,\theta)
    +f^{\rm RP}(\omega,\theta),
\end{equation}
where the background contribution is
\begin{equation}
\label{eq:CAM_background_decomposition}
f^{\rm B}(\omega,\theta)=f^{{\rm B},{\rm Re}}(\omega,\theta)+f^{{\rm B},{\rm Im}}(\omega,\theta),\end{equation}with\begin{align}f^{{\rm B},{\rm Re}}(\omega,\theta)&=\frac{1}{\pi\omega}\int_0^\infty d\lambda\,\lambda S_{\lambda-1/2}(\omega)Q_{\lambda-1/2}(\cos\theta+i0),
\label{eq:cam_background_real}\\
f^{{\rm B},{\rm Im}}(\omega,\theta)&=\frac{1}{\pi\omega}\int_0^\infty dy\,y S_{iy-1/2}(\omega)Q_{iy-1/2}(\cos\theta+i0).
\label{eq:cam_background_imag}
\end{align}
Here $Q_{\nu}$ is the Legendre function of the second kind. The two terms arise, respectively, from the real and imaginary CAM contours and contain the nonresonant part of the amplitude. The pole contribution is
\begin{equation}\label{eq:RP_amplitude}
    f^{\rm RP}(\omega,\theta)
    =-\frac{i\pi}{\omega}
    \sum_{n=0}^{\infty}
    \frac{\lambda_n(\omega)r_n(\omega)}
         {\cos\!\left[\pi\lambda_n(\omega)\right]}
    P_{\lambda_n(\omega)-1/2}(-\cos\theta).
\end{equation}
\eqref{eq:CAM_decomposition}--\eqref{eq:RP_amplitude} provide a representation of the scattering amplitude. 

\subsection{Regge pole shifts}
We determine the RPs by analytically continuing \eqref{radialeqn} to complex angular momentum through  $\ell=\lambda-1/2$. For each real frequency $\omega$, one independent solution is integrated outward from the horizon with a purely ingoing boundary condition, while a second solution is integrated inward from a sufficiently large radius using the asymptotic behaviour determined by the ADM mass, $M_{\rm ADM}=M_{\rm BH}+M$. The two solutions are matched at an intermediate radius. An RP is identified when the coefficient of the incoming wave at infinity vanishes, which is equivalently characterised by the vanishing of the matching Wronskian. This numerical procedure closely follows the algorithm of \cite{Pezzella:2024tkf}, except that the angular momentum $\lambda$ is treated as the eigenvalue while the frequency $\omega$ is held fixed and real.

\begin{figure}[t]
    \centering
    \includegraphics[width=\columnwidth]{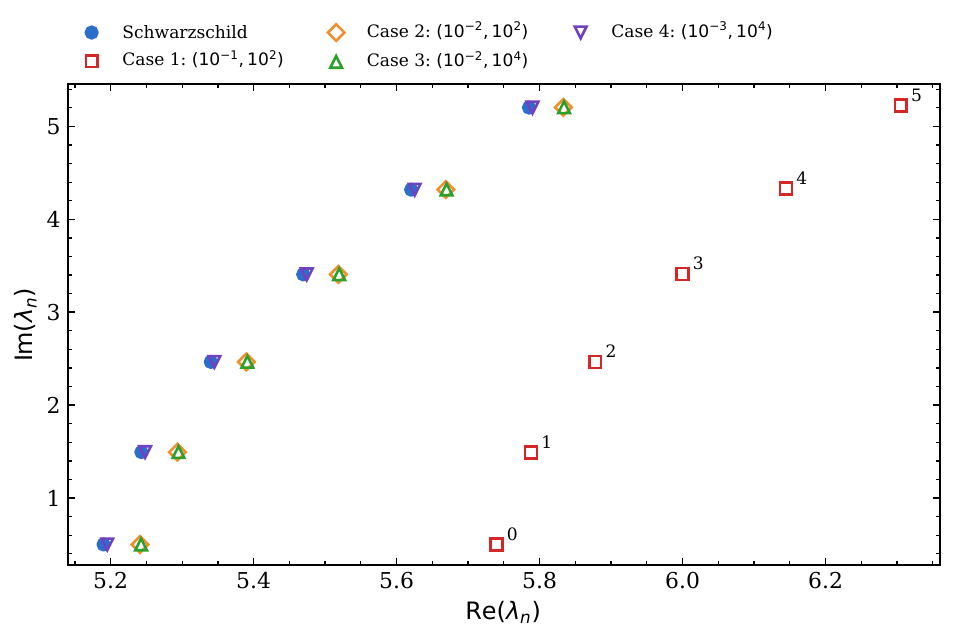}
    \caption{ First six scalar Regge poles at $M_{\rm BH}=1$ and $\omega=1$. Blue circles denote Schwarzschild, red squares, orange diamonds, green upward triangles, and purple downward triangles denote Cases~1--4, respectively. The integers label the Case~1 overtones $n=0,\ldots,5$.} 
    \label{fig:diff_z_w}
\end{figure}

Fig.~\ref{fig:diff_z_w} compares the first six RPs of the four cases \eqref{eq:benchmark_cases} with the Schwarzschild trajectory at $\omega=1$ and $M_{\rm BH}=1$.
The displacement is dominated by $\operatorname{Re}\lambda_n$, whereas $\operatorname{Im}\lambda_n$ changes only weakly for low overtones. The halo therefore shifts the angular interference pattern while producing a smaller change in the damping of the orbiting contributions. This provides a spectral interpretation of Fig.~\ref{fig:glory2}.
Together, the geodesic, partial wave, and RP results give a consistent description of the halo induced modification of wave propagation near the unstable null orbit.

To quantify the halo induced shift of the RPs, we define \begin{equation}\label{eq:rp_spectral_displacement}
    \Delta^{\lambda}_n(z,a_0,\omega)
    \equiv
    \frac{\left|\lambda^{\rm halo}_n(z,a_0,\omega)
    -\lambda^{\rm Schw}_n(\omega)\right|}
    {\left|\lambda^{\rm Schw}_n(\omega)\right|}
\end{equation}
which measures the relative pole displacement and allows comparisons across overtone number, halo compactness, and scale radius. Its variation with $n$ distinguishes an overtone dependent reorganisation from a coherent displacement of the pole sequence. 
We use the first three RPs to compare the four benchmark configurations at $M_{\rm BH}=1$ and $\omega=1$. The first six RPs are used below for the CAM reconstruction. 

\begin{table}[t]
    \centering
    \caption{Relative displacements $\Delta_n^\lambda$ of the first three RPs, $n=0,1,2$, from their Schwarzschild values at $M_{\rm BH}=1$ and $\omega=1$. The four halo configurations are defined in \eqref{eq:benchmark_cases}.}    \label{tab:rp_displacements}
    \begin{tabular}{lccc}
        \toprule
        Configuration & $\Delta_0^{\lambda}$ & $\Delta_1^{\lambda}$ & $\Delta_2^{\lambda}$ \\
        \midrule
        Case 1 & $1.0539\times10^{-1}$ & $9.9954\times10^{-2}$ & $9.1435\times10^{-2}$ \\
        Case 2 & $9.7673\times10^{-3}$ & $9.2564\times10^{-3}$ & $8.4606\times10^{-3}$ \\
        Case 3 & $1.0057\times10^{-2}$ & $9.5240\times10^{-3}$ & $8.6932\times10^{-3}$ \\
        Case 4 & $9.9817\times10^{-4}$ & $9.4518\times10^{-4}$ & $8.6264\times10^{-4}$ \\
        \bottomrule
    \end{tabular}
\end{table}
For reference, the Schwarzschild fundamental pole is $\lambda_0^{\rm Schw}=5.190075+0.501235i$, whereas Case~1 gives $\lambda_0=5.739620+0.499608i$. The displacement is dominated by $\operatorname{Re}\lambda_0$, while $\operatorname{Im}\lambda_0$ changes only weakly. Table~\ref{tab:rp_displacements} shows an approximately tenfold reduction in $\Delta_n^{\lambda}$ when $z$ decreases by one order of magnitude at fixed $a_0$. At fixed $z=10^{-2}$, changing $a_0$ from $10^4$ to $10^2$ changes $\Delta_n^{\lambda}$ by only a few percent. Compactness is thus, again, the leading control parameter, while the scale radius gives a secondary correction. The displacement decreases mildly from $n=0$ to $n=2$.

\subsection{Scattering cross section reconstruction} 
\begin{figure*}[t]
    \centering
    \includegraphics[width=\textwidth]{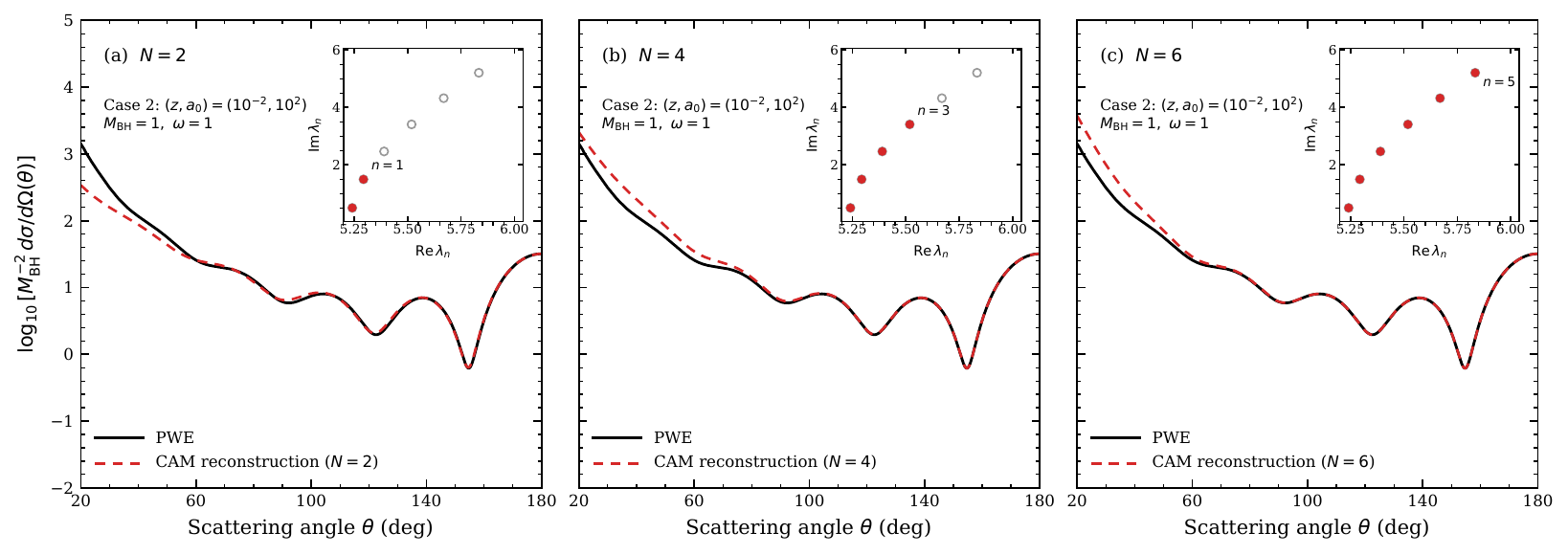}
    \caption{ Partial wave result and CAM reconstructions of the scalar differential cross section for the Case~2 halo, $(z,a_0)=(10^{-2},10^2)$, at $M_{\rm BH}=1$ and $\omega=1$. The vertical axis shows $\log_{10}[M_{\rm BH}^{-2}d\sigma/d\Omega]$, as in Fig.~\ref{fig:glory2}. Each CAM curve includes both background contours and $N=2$, $4$, or $6$ RPs.}
    \label{fig:cam_case2_reconstruction}
\end{figure*}

To verify that the CAM representation also reproduces the scattering cross section, we define the truncated amplitude
\begin{equation}\label{eq:cam_truncated_amplitude}
    f^{\rm CAM}_N(\omega,\theta)
    \equiv f^{\rm B}(\omega,\theta)
    +\sum_{n=0}^{N-1} f^{\rm RP}_n(\omega,\theta),
\end{equation}
The two CAM background contours in \eqref{eq:cam_background_real} and \eqref{eq:cam_background_imag} are evaluated independently, rather than inferred by subtracting the pole sum from the partial wave result. We then introduce the normalised reconstruction error
\begin{equation}\label{eq:cam_reconstruction_error}
    \mathcal{E}_N(\omega,\theta_{\min})
    \equiv
    \left[
    \frac{
    \displaystyle\int_{\theta_{\min}}^{\pi}
    \left|f^{\rm PWE}(\omega,\theta)
    -f^{\rm CAM}_N(\omega,\theta)\right|^2
    \sin\theta\,d\theta}
    {\displaystyle\int_{\theta_{\min}}^{\pi}
    \left|f^{\rm PWE}(\omega,\theta)\right|^2
    \sin\theta\,d\theta}
    \right]^{1/2},
\end{equation}
where $f^{\rm PWE}$ is the amplitude obtained from the exact partial wave sum. The decrease of $\mathcal E_N$ with $N$ directly measures how rapidly the pole spectrum reconstructs the observable after the CAM background has been included.

We evaluate the pole positions, residues, and background contours defined above\footnote{The slowly convergent real axis integral is treated with the continuous version of the series reduction used in \eqref{eq:yrw_amp} and \eqref{eq:yrw_recursion}, while the imaginary axis integrand decays exponentially.}.
We perform the complete reconstruction for the Case~2  configuration, $(z,a_0)=(10^{-2},10^2)$, at $M_{\rm BH}=1$ and $\omega=1$. The first six poles and residues are listed in Table~\ref{tab:cam_case2_residues}. Their normalisation is validated against published Schwarzschild results \cite{Folacci:2019cmc}, and the final reconstruction is stable under variations of the numerical boundaries.

\begin{table}[t]
    \centering
    \caption{First six scalar RPs and residues for the Case~2 Hernquist halo at $M_{\rm BH}=1$ and $\omega=1$.}
    \label{tab:cam_case2_residues}
    \begin{tabular}{ccc}
        \toprule
        $n$ & $\lambda_n$ & $r_n$ \\
        \midrule
        0 & $5.241004+0.501072i$ & $-0.075344+0.746004i$ \\
        1 & $5.293459+1.494602i$ & $ 2.563489+0.853095i$ \\
        2 & $5.389854+2.465823i$ & $ 3.764254-3.736377i$ \\
        3 & $5.518452+3.408120i$ & $-0.883127-7.812066i$ \\
        4 & $5.668818+4.320480i$ & $-7.605369-6.561836i$ \\
        5 & $5.833248+5.204768i$ & $-11.759532-0.986668i$ \\
        \bottomrule
    \end{tabular}
\end{table}

Fig.~\ref{fig:cam_case2_reconstruction} shows systematic convergence of the CAM reconstruction to the partial wave result as the number of poles increases. With six poles, the normalised reconstruction mismatch is $0.32\%$ over $80^\circ\leq\theta\leq180^\circ$, $0.12\%$ over $90^\circ\leq\theta\leq180^\circ$, and $0.01\%$ over $120^\circ\leq\theta\leq180^\circ$. The backward scattering and glory sectors are therefore reconstructed with high accuracy. 

\section{Conclusion}\label{sec:conclusion}
We have studied the scattering of massless scalar waves by a static and spherically symmetric black hole surrounded by a dark matter halo using null geodesics, the glory approximation, the partial wave expansion, and the complex angular momentum method. The Hernquist and Navarro--Frenk--White metric functions show qualitatively similar behaviour for the configuration considered,  hence we focused the scattering analysis on the exact Hernquist geometry. 

\par At the geodesic level, the halo only weakly shifts the light ring radius but produces more pronounced shifts in the critical and glory impact parameters relative to their Schwarzschild values. Near the backward direction, the glory approximation reproduces the characteristic oscillatory structure of the partial wave result. Together, these results provide a qualitative semiclassical interpretation of the halo induced modification of backward scattering. The full partial wave cross section confirms that the characteristic black hole interference pattern is retained. Within the parameter space explored, the halo compactness provides the leading correction and the scale radius produces a secondary effect. For the most compact configuration $(z,a_0) = (10^{-1}, 10^2)$, a redshift-corrected Schwarzschild frequency $\omega_{\rm eff}\simeq1.12$ reproduces most of the displacement of the interference extrema. The corresponding solid angle weighted global residual is $9.51\%$, demonstrating that the halo scattering pattern is not fully degenerate with a frequency shift. 
\par At fixed $\omega=1$, the halo induces a coherent shift of the Regge poles, predominantly in $\operatorname{Re}\lambda_n$, with much smaller changes in their imaginary parts. Within the parameter range explored, no runaway behaviour or additional branch is observed in the resolved spectrum. For the representative configuration $(z,a_0)=(10^{-2},10^2)$, we combine the first six RPs and their residues with both independently evaluated CAM background contours to reconstruct the large angle scattering amplitude with subpercent accuracy. This provides a full CAM reconstruction of the scattering cross section in an exact, smooth halo geometry. 
\par The present analysis is restricted to a static and spherically symmetric background and a scalar probe. Extensions to fields with spin, rotating black holes, and more realistic matter distributions are required for observational applications. Such studies should retain the full CAM reconstruction because displacement of individual poles alone does not necessarily translate into an observable environmental signature.

\begin{acknowledgments}
We would like to thank Mohamed Ould El Hadj, Xiao-Mei Kuang, Zhen-Hao Yang, and Filippo Vernizzi for helpful discussions and useful suggestions on this work.
This work received support from the French government under the France 2030 investment plan, as part of the Initiative d'Excellence d'Aix-Marseille Universit\'e - A*MIDEX (AMX-19-IET-012). It was also supported by the ``action th\'ematique" Cosmology-Galaxies (ATCG) of the CNRS/INSU PN Astro and by the {\it Agence Nationale de la Recherche} under the grant ANR-24-CE31-6963-01.
\end{acknowledgments}

\bibliographystyle{apsrev4-1}
\bibliography{ref}

\end{document}